\documentclass[aps,pra,superscriptaddress
,12pt,tightenlines,final,
twoside,showpacs]{revtex4}

\usepackage{graphicx}
\usepackage{color}
\usepackage{epsfig}
\usepackage{amsmath}
\usepackage{amstext}
\usepackage{amssymb}
\usepackage{bm}
 \DeclareGraphicsExtensions{.eps,.ps}

\begin{document}

\author{Ravindra W. Chhajlany }

\author{Piotr Tomczak}

\author{Antoni W\'ojcik}

\affiliation
{
Physics Department, Adam Mickiewicz University,
Umultowska 85, 61-614 Pozna\'n, Poland
}

\author{Johannes Richter}

\affiliation{Institut f\"ur Theoretische Physik, Universit\"at Magdeburg,
      P.O. Box 4120, D-39016 Magdeburg, Germany}

\begin{abstract}

  We present an analysis of the entanglement characteristics in the
  Majumdar-Ghosh (MG) or $J_{1}$-$J_{2}$ Heisenberg model. For a
  system consisting of up to 28 spins, there is a deviation from the
  scaling behaviour of the entanglement entropy characterizing the
  unfrustrated Heisenberg chain as soon as $J_{2} >0.25$.  This
  feature can be used as an indicator of the dimer phase transition
  that occurs at $J_{2} = J_{2}^{*} \approx 0.2411 J_{1}$.
  Additionally, we also consider entanglement at the MG point
  $J_{2}=0.5 J_{1}$.

\end{abstract}

\pacs{
03.67.Mn,       
05.70.Jk,       
75.10.Jm
}

\date{\today}
\title{Entanglement in the Majumdar-Ghosh model}

\maketitle

\section{Introduction}\label{}

Entanglement has come to be seen as a prime resource for various
quantum information processing tasks \cite{nc}.  From another angle,
entanglement describes quantum correlations of many body systems which
on their part are responsible for various interesting physical
phenomena, {\it e.g.} quantum phase transitions. In recent
years, entanglement has been fruitfully used to give an alternate view
on quantum phase transitions, especially in low dimensional quantum systems (see
\cite{Vidal03,Roscilde} and references therein).  Frustrated quantum
systems, have however been largely left out of the picture in such
discussions so far.

In this Article, we consider the entanglement in a prototypical 1-D
frustrated spin system -- the Majumdar Ghosh or $J_{1}$-$J_{2}$  Heisenberg
chain \cite{mg}. The spin-1/2 Hamiltonian is described by   
\begin{gather}
      H = J_1  \sum _{\langle n.n.\rangle}{\bf S}_i \cdot {\bf S}_j 
       + J_2  \sum _{\langle n.n.n. \rangle}{\bf S}_i \cdot {\bf S}_j,
\label{R43a}
\end{gather}
where the sums run over nearest neighbour (n.n.)  and next nearest
neighbour (n.n.n.) spins, and ${\bf S}_{i}$ are spin-1/2 operators. In
what follows, we set $J_{1}=1$ as the energy scale and consider
antiferromagnetic $J_{2} >0$ interaction. The ground state properties
of this system have been studied so far with the use of many methods
like exact diagonalization, DMRG, field theoretical approach (see
\cite{White} and \cite{Mikeska} for an overview). It is known that the
model is critical (i.e., the spin-spin correlation function decays
algebraically) and gapless for $J_2 \le J^*_2 \approx 0.2411 $
\cite{Mikeska,Okamoto}.  A gap to triplet excitations opens beyond the
quantum critical point $J^*_2 $ accompanied also by the stabilization
of a dimerized phase.  Interestingly, for $J_{2}=0.5$ (known as the MG
point), the ground state manifold is exactly known. It is spanned
by two degenerate nearest neighbour dimer states given by
\begin{gather}
|R \rangle = (1,2)(3,4)\ldots (N-1,N) ,\; \; \; |L \rangle =
 (2,3)(4,5) \ldots (N,1)
\label{R53a}
\end{gather}
where $(i,j)$ denotes a singlet between spins $i$ and $j$. 

\begin{figure}[t] 
\centerline{\resizebox{1.0\columnwidth}{!}{\rotatebox{0}
{\includegraphics{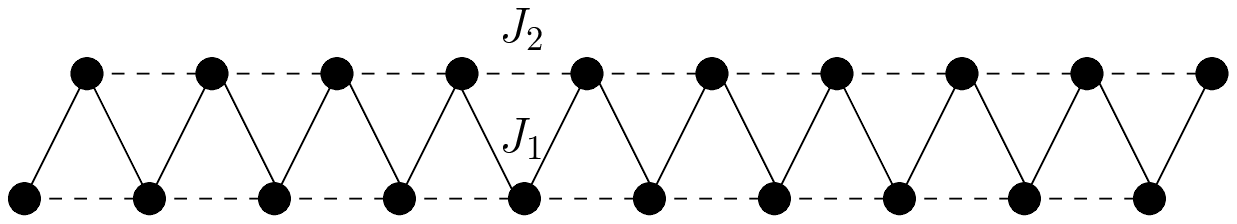}}}}
\caption{The model system with nearest $J_{1}$ and next nearest neighbour
  interaction $J_{2}$}
\label{fig1}
\end{figure}

In the following, we will demonstrate that the entanglement, as
quantified by the entanglement entropy, exhibits characteristic
scaling properties in the critical region.  We will show that the
ground state entanglement scales not only with respect to the
subsystem size $n$ in an infinite system (this possibility was
examined carefully in case of models which can be solved exactly
\cite{Vidal03}), but in finite systems for fixed $n$ also with respect
to the total size $N$ of the system. In the gapped phase, a clear
deviation from this scaling emerges which can be used to identify the
critical point of this system. This is to be contrasted with the fact
that the entanglement measured by standard measures such as
entanglement entropy or pair-wise concurrence do not detect the
critical point of this model, on their own.

In the second part of this paper, we consider the entanglement at the
MG point. This has been considered earlier in Ref.
\onlinecite{qian}, where it was shown that the nearest neighbour
concurrence exhibits a jump at the MG point in finite systems.
However this jump disappears exponentially as the total size of the
system increases. We derive a simple formula for the change of
concurrence as the parameter $J_{2}$ is made to pass through the MG
point. Furthermore, we consider a different quantity, the entanglement
of a pair of next-nearest neighbour spins, as a potential indicator of
the MG point. This quantity is maximal at the MG point for any size of
the system. Furthermore, we analytically show that in the
thermodynamic limit this entanglement entropy is invariant in the MG
manifold, and thus is a robust maximum.  Interestingly as a byproduct,
in the small size limit, the two translationally invariant (or
'quasi-momentum') states turn out to have entanglement entropies that
are local minima in this manifold.

\section{Entanglement entropy scaling}\label{}

Given a pure state of a quantum many body system, the entanglement of
a given subsystem with its complement is conveniently measured by the
entanglement entropy. The entanglement entropy is defined as the von
Neumann entropy of the chosen subsystem, {\it i.e.} if the (reduced)
density matrix of the subsystem is $\rho_{n}$ and $\lambda_{j}$ are
its eigenvalues, then the entanglement entropy is defined as
\begin{gather}
S(\rho_{n}) \equiv - \sum_{j}^{} \lambda_{j} \log_{2} \lambda_{j}.
\label{R63a}
\end{gather}

The entanglement entropy of a block of contiguous spins has been shown
to scale differently with the block size $n$, in critical and
non-critical 1D systems \cite{Vidal03}. In the thermodynamic limit
(total system size $N \rightarrow \infty $), the entanglement entropy
of a non-critical system tends to saturate, while it displays
universal scaling behaviour for critical systems \cite{Holzhey}:
\begin{gather}
S(n) = c_{0} + \frac{c}{3} \log_{2}n,
\label{R73a}
\end{gather}
where $c$ is a universal scaling constant and $c_{0}$ is model
dependent. These two constants take the values $1$ and $\pi $ for the
isotropic Heisenberg antiferromagnetic chain.
The extension of Eq.(\ref{R73a}) to finite critical systems was given
in \cite{Calabrese}:
\begin{gather}
S(n,N)= c_{0} +\frac{c}{3} \log_{2}\Big(\frac{N}{\pi } \sin \big(\pi
\frac{n}{N}\big)\Big).
\label{R93a}
\end{gather}
For small values of $n/N$, this equation becomes
\begin{gather}
S(n,N) = c_{0} +\frac{c}{3} \log_{2}n - \frac{c}{18 \ln2} \pi^{2}
\Big(\frac{n}{N}\Big)^{2} - \frac{c}{540 \ln 2}
\pi^{4}\Big(\frac{n}{N}\Big)^{4} + {\cal O}\Big(\frac{n}{N}\Big)^{6}
\label{R03a}
\end{gather}

In the rest of this section, we consider the scaling of the block
entanglement entropy in the ground state of the MG model. The ground
state is calculated via exact diagonalization (Lanczos method) of
systems of up to 28 spins on imposing periodic boundary conditions.

\subsection{Case 1: Fixed $N=28$}\label{}

\begin{figure}[t] 
\centerline{\resizebox{0.5\textheight}{!}{\rotatebox{270}
{\includegraphics{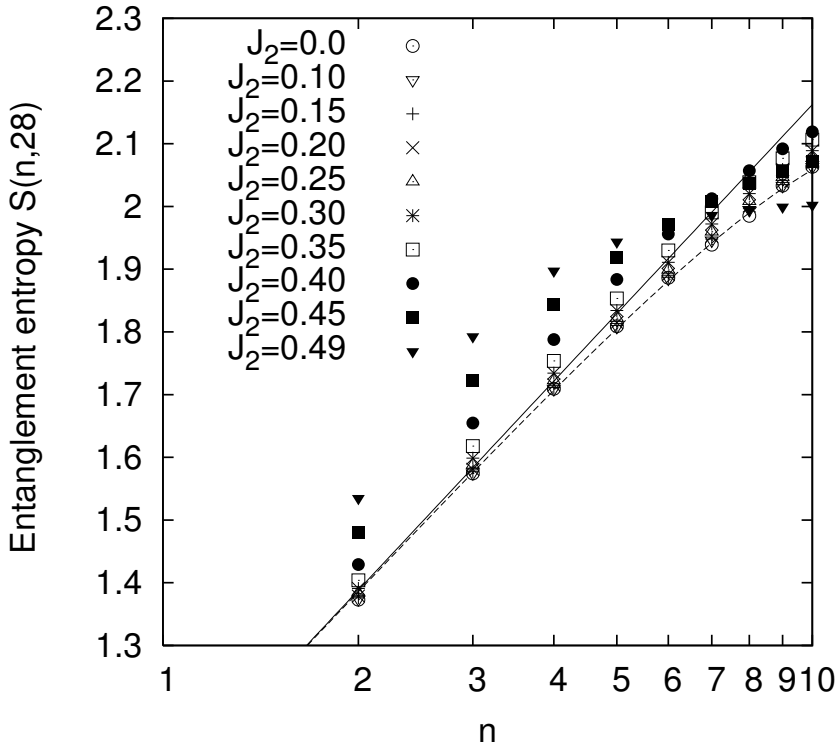}}}}
\caption{The scaling of the entanglement entropy of the MG model.  The
system contains 28 spins and the length of the subsystem changes from
$n=2$ to $n=10$.  The full line represents the logarithmic scaling of an
infinite system --- Eq. (\ref{R73a}). The dashed line shows the finite
size correction given by Eq. (\ref{R93a}). Note that the correction
resulting due to $J_2 > 0$ is most pronounced for small $n$.  For
larger subsystem sizes the finite size correction dominates. }
\label{fig2}
\end{figure}

First, let us focus on the scaling of the entanglement entropy w.r.t.
the block size $n$ of consecutive spins for a system with a fixed
number $N$ of spins.  This case, for $N \le 20$ and $J_2 = 0$, has
been analyzed previously \cite{Vidal03}. Our results for $N=28$ are
shown in Fig. \ref{fig2}.  As expected the numerical data for $J_2=0$
(open circles) are well described by Eqs.  (\ref{R93a}) or
(\ref{R03a}) describing the saturation of von Neumann entropy for
finite $N$.  But there is also good agreement of the calculated
entanglement for finite $J_2$ up to $J_2 \approx 0.25$ with the line
given by Eq. (\ref{R93a}).  Fitting the values of $S(n,28)$ to Eq.
(\ref{R93a}) yields $c_0=3.131$ and $c=1.017$.  For comparison, the
logarithmic divergence of the entanglement in the thermodynamic limit
is also drawn.

The correction to this scaling due to the frustrating $J_2$ in the
finite system of 28 spins can be most clearly seen for small $n$. For
larger $n$ the finite-size correction (see Eq. (\ref{R03a})) and the
frustration effect show opposite tendencies and cancel each other
partially.  However, for strong frustration near the Majumdar-Gosh
point, i.e. for $J_2=0.49$, the frustration effect is clearly visible
for all $n$ considered.  It is thus reasonable to argue that the
presence of $J_2 \ne 0$ will produce a saturation of the entanglement
vs. $n$.

\begin{figure}[t] 
\centerline{\resizebox{0.5\textheight}{!}{\rotatebox{270}
{\includegraphics{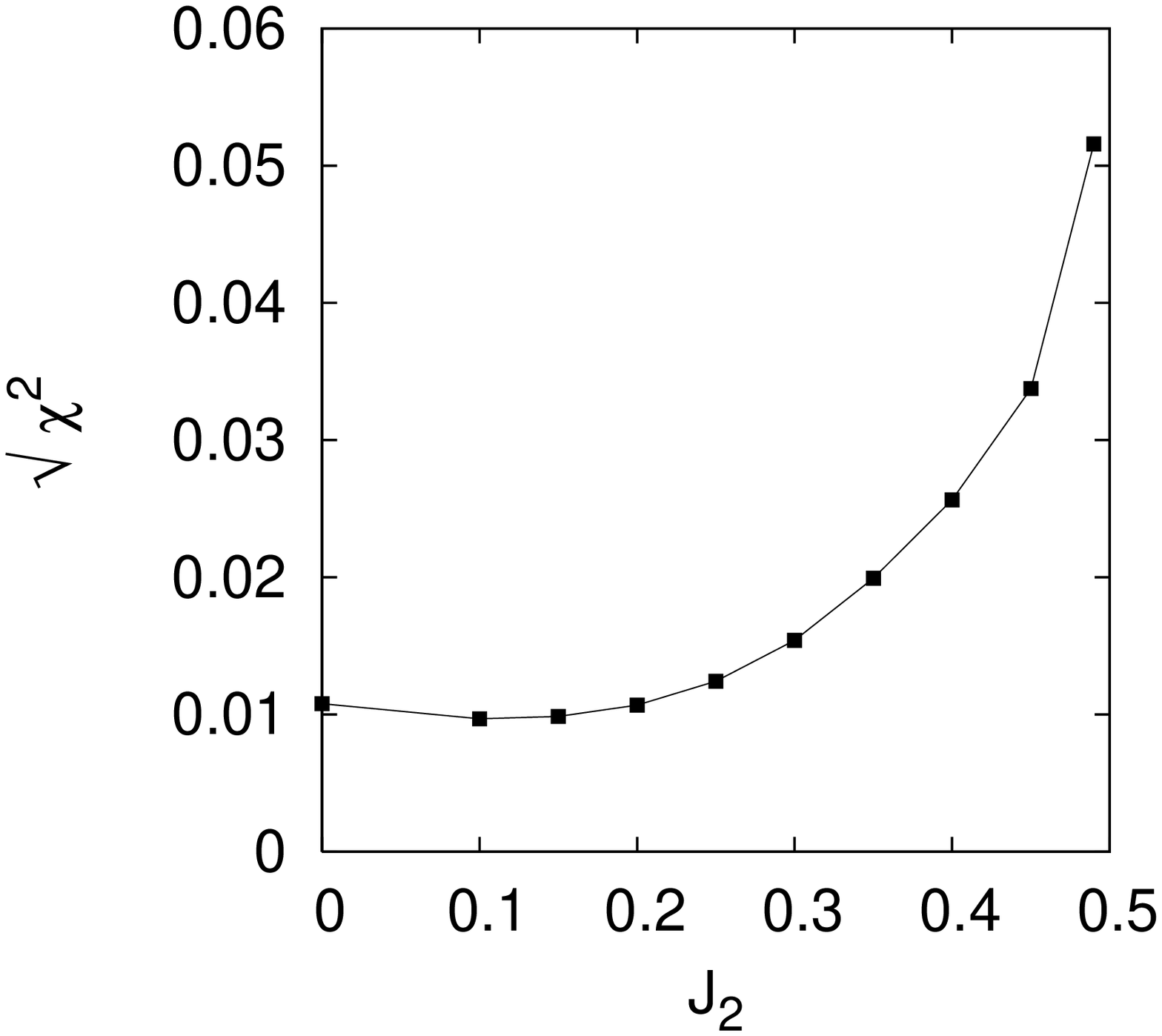}}}}
\caption{The least-square deviation $\sqrt{\chi^2}$ of the calculated
values of the entanglement form the line described by Eq. \ref{R93a}
(dashed line in Fig. {\ref{fig2}}) in dependence on the frustration
$J_2$.  The system contains 28 spins and $\chi^2$ was calculated for
$n \le 6$.  Note that $\sqrt{\chi^2}$ starts to increase as beyond the
quantum critical point $J^*_2 \approx 0.2411$.}
\label{fig3}
\end{figure}

In order to quantitatively characterize the deviation of the
entanglement from the scaling relation (\ref{R93a}) at finite
frustrating $J_2>0$, the value of $\chi^2$ defined in the standard
way, i.e., as a square root of the sum of squares of residuals, is
calculated and plotted in Fig. \ref{fig3}.  This quantity measures the
differences between the calculated entropies for $J_2>0$ and the line
given by Eq. (\ref{R93a}). Notably,  one can observe a significant
deviation from the critical scaling for $J_2 \gtrsim J^*_2 \approx
0.2411$.  The flat minimum for $J_2 \approx 0.10$ is a result of the
competition between a finite size correction and the correction
resulting from the presence of interaction $J_2$.

\subsection{Case 2: Fixed $n$}\label{}

The question arises as to whether the quantum phase transition in the
MG model can be identified directly from the dependence of
``sufficiently local'' entanglement measures on the control parameter
$J_{2}$, as in certain other models. Indeed, as shown by Osterloh {\it
et al.}  \cite{osterloh}, the phase transitions in XY spin-1/2 models
can be identified explicitly by the dependence of the entanglement
between two nearest neighbour spins, as measured by the concurrence,
on the control parameter. On the other hand, single and two-spin
entanglement entropy of XY models, have been shown to detect
quantum phase transitions (see \cite{venuti,nielsen,su}).

The concurrence does not visually detect the dimer phase transition of
the MG model, as seen in Ref. \onlinecite{qian}.  The single spin
entropy is always equal to $1$, since the ground state remains
rotationally invariant (see Section \ref{sec2}).  We have checked that
entanglement entropies of variously chosen subsystems, such as block
entanglement of consecutive spins, or blocks of next nearest neighbour
spins also do not identify the phase transition in this model.  The
reason for this may be that: either local entanglement measures do not
detect this phase transition or the small system sizes considered do
not capture this behaviour \footnote{The systems sizes studied in Ref.
  \onlinecite{qian} range from $N=6$ to $12$, while we have densely
  checked for up to $N=18$.}. 

\begin{figure}[t] 
\centerline{\resizebox{0.5\textheight}{!}{\rotatebox{270}
{\includegraphics{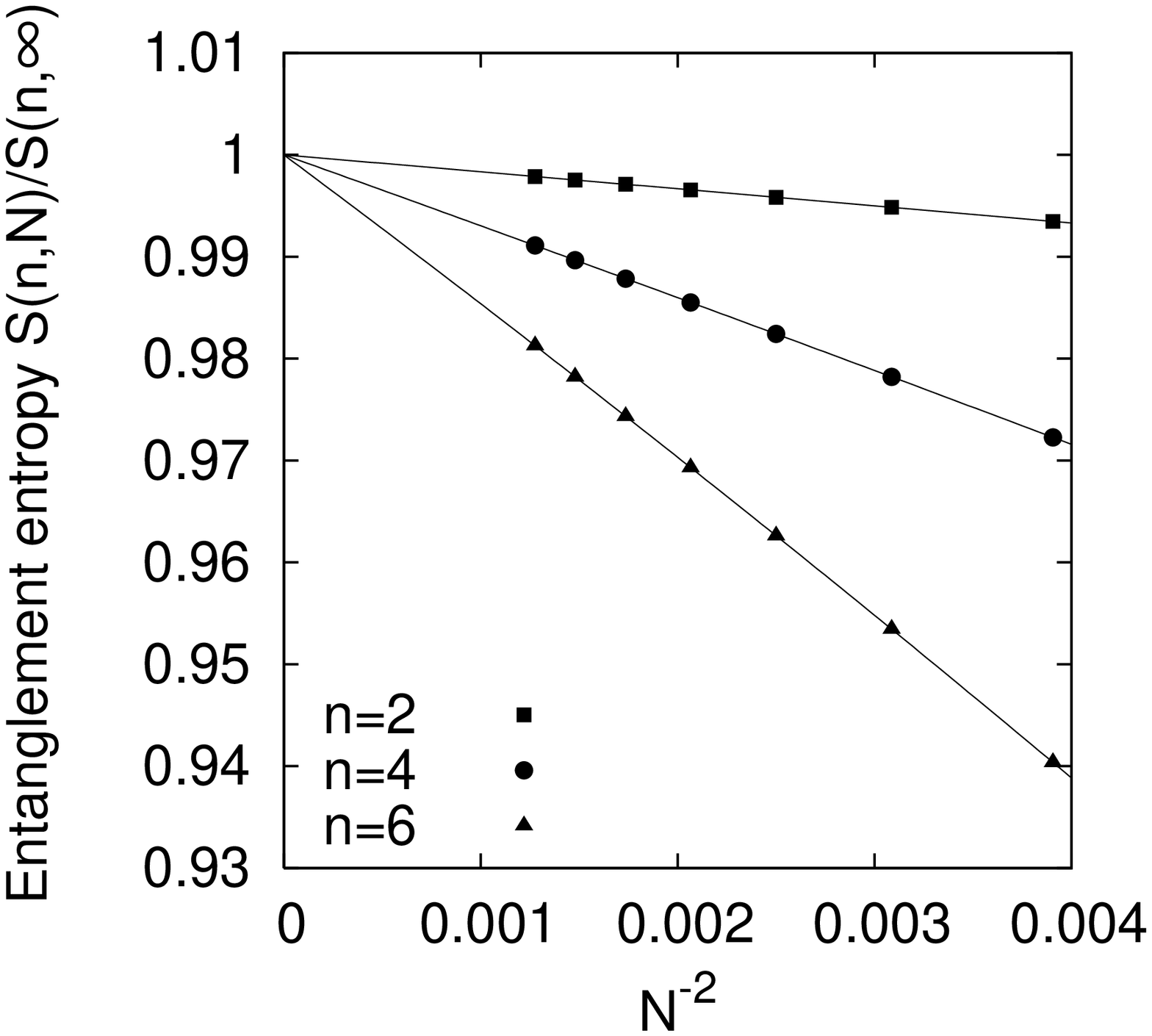}}}}
\caption{Finite-size scaling of the von Neumann entropy $S(n,N)$ of
the unfrustrated Heisenberg chain, i.e. for $J_2=0$, for fixed
subsystem size n.  The system size $N$ changes from 16 to 28 spins.}
\label{fig4}
\end{figure}

The finite size problem can be circumvented by the usual finite-size
scaling techniques. Consider {\it e.g.} the entanglement entropy of
two nearest neighbour spins ($n=2$) for different total system sizes
$N$. For the isotropic Heisenberg antiferromagnet ($J_{2}=0$)  this quantity scales as
$N^{-2}$ as shown in Fig. \ref{fig4} (the scaling is given by a
perfect line with correlation better than 0.9999 for N ranging from 16
to 28). This fully agrees with the dominating behaviour in
Eq.(\ref{R03a}). The scaling of blocks of $4,6$ spins are also
presented in Fig. \ref{fig4} \footnote{For these and larger blocks,
effects of terms of the order ${\cal O}$ $(n/N)^{4}$ become
important.}.

\begin{figure}[t] 
\centerline{\resizebox{0.5\textheight}{!}{\rotatebox{270}
{\includegraphics{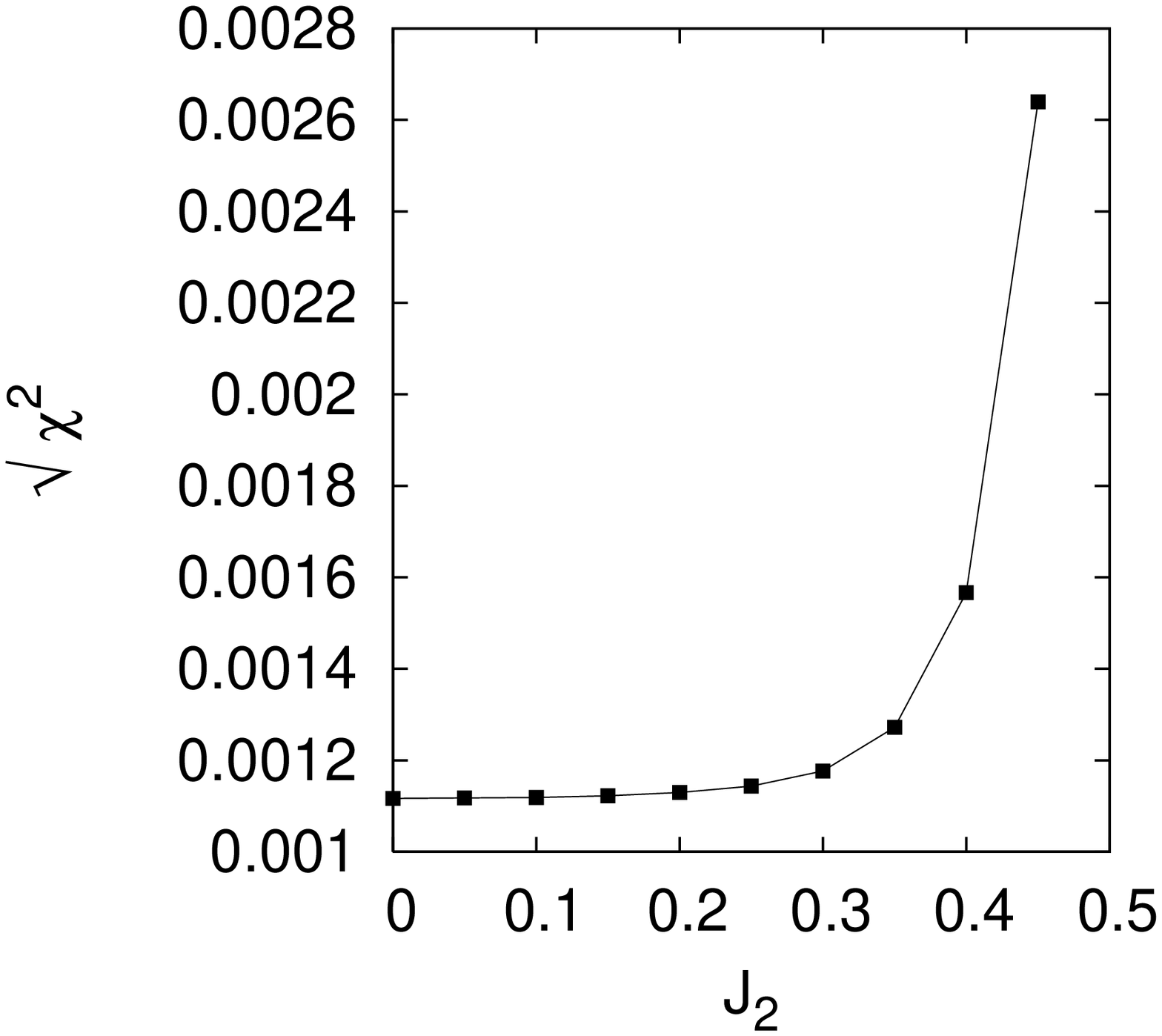}}}}
\caption{The deviation (measured by $\sqrt{\chi^2}$) from the straight
line scaling for $J_2=0$ and $n=2$ (the upper line in Fig. \ref{fig4})
for bigger values $J_2$.}
\label{fig5}
\end{figure}

Let us focus on the dependence of the scaling of $n=2$ entanglement
entropy on the control parameter $J_{2}$. The correction to the
$N^{-2}$ scaling can again be characterized by the value of
$\chi^{2}$, describing the difference between the calculated entropies
for $J_{2}>0$ and the straight line for $J_{2}=0$, see
Fig. \ref{fig5}. Significantly, there is a clear deviation from
critical scaling for $J_{2}>J_{2}^{*}$. The phase transition may thus
be detected and located by the dependence of $\sqrt{\chi }^{2}$ on
the control parameter. 

While the above results show that there is a change in scaling of
local entanglement entropy around the critical point, the general
question posed in this section remains open for future discussion.

\section{The Majumdar-Ghosh point}\label{sec2}

We now turn to the entanglement at the MG point $J_{2}=1/2$.  In
particular, we are going to consider two measures of entanglement {\it
  viz}. the concurrence of two spins (which was recently analyzed in
\cite{qian}) and the entanglement entropy of two spins.

To fix notions, recall that the model Hamiltonian (Eq.(\ref{R43a}))
possesses rotational ($SU(2)$) symmetry. It is widely believed that
the ground state $|g \rangle $ of 1-dimensional Heisenberg
antiferromagnets also exhibits this symmetry, {\it i.e.} is a total
singlet $S=0$ state. This implies that any subset of spins chosen from
the whole system is also rotationally invariant. This follows from the
following simple identity for the reduced state of $n$ arbitrary spins:
\begin{gather}
\rho_{n} = \text{Tr}'\rho_{g} = \text{Tr}'U^{\otimes N} \rho_{g}(U
^{\dagger})^{\otimes N} = U^{\otimes n} \big(\text{Tr}'U^{\otimes N-n}
\rho_{g}(U ^{\dagger})^{\otimes N-n}\big)(U ^{\dagger})^{\otimes n} =
U^{\otimes n} \rho_{n} (U ^{\dagger})^{\otimes n},
\label{Ra4a}
\end{gather}
which holds provided $\rho_{g}= |g \rangle \langle g|$ is rotationally
invariant, {\it i.e.}  $\rho_{g}=U^{\otimes N} \rho_{g}\; (U
^{\dagger})^{\otimes N} $ for an arbitrary single spin unitary
operator $U$ (the symbol ${\rm Tr}'$ denotes the partial trace over the
unwanted spins).  In particular, the reduced state of each individual
spin in the ground state of the MG model is maximally mixed $\rho_{1}
= \openone/2$ implying that a single spin is maximally entangled with
either all or some of the remaining spins in the lattice. Similarly,
any state of two spins belongs to the one parameter family of
so-called Werner states \cite{werner}, that can be represented as
\begin{gather}
\rho_{2} = p |\Psi^{-}\rangle \langle \Psi^{-}|+ \frac{1-p}{4} \openone,
\label{Rb4a}
\end{gather}
where $|\Psi_{-}\rangle $ denotes a singlet and $p=-(4/3) \langle
\bm{S}_{i} \cdot \bm{S}_{j}\rangle $ is just the (rescaled) isotropic
correlation function of the involved spins.

The entanglement between two spins, characterized by the concurrence
\cite{wootters} $C(\rho_{2})$, can be checked to be given by a simple
formula for Werner states
\begin{gather}
C(\rho_{2}) = \max \Big(0, \frac{3}{2}p - \frac{1}{2}\Big).
\label{Rc4a}
\end{gather}
Thus, two spins are entangled with each other as long as the
correlations between them are sufficiently antiferromagnetic,
$\langle\bm{S}_{i} \cdot \bm{S}_{j} \rangle < -1/4$
\footnote{Interestingly, $\langle\bm{S}_{i} \cdot \bm{S}_{j} \rangle =
  -1/4$ is the classically optimal antiferromagnetic correlation
  corresponding to the N\'eel state. So, enhanced antiferromagnetic
  correlations are a manifestation of quantum entanglement.}. The
entanglement entropy (see Eq. (\ref{R63a})) of two spins in a Werner
state, on the other hand, is given by the relation
\begin{gather}
S(\rho_{2})=2 - \frac{1+3p}{4} \log (1+3p) - 3\frac{1-p}{4} \log(1-p).
\label{Rd4a}
\end{gather}

Returning now to the MG point, one of the states $|R \rangle $ or $|L
\rangle $ is realized as the ground state, with broken translation
symmetry, in the thermodynamic limit. In these states, the nearest
neighbour concurrence is either $0$ or $1$ depending on whether the
considered pair resides on the same or different singlets. The average
nearest neighbour concurrence in both these states is $C(|R(L) \rangle
)_{av} = 1/2$. For finite chains however, in the absence of an
additional symmetry breaking field, it is more natural to revert to an
orthogonal ``qubit'' basis of the ground state manifold, which can be
chosen to be the eigenstates of the momentum operator
\begin{gather}
|\pm \rangle = \frac{1}{\sqrt{\Omega_{\pm}}}(|R \rangle \pm |L \rangle
 ),
\label{Re4a}
\end{gather}
where the normalizing factors are 
\begin{gather}
\Omega_{\pm} = 2(1 \pm x), \; \; x \equiv \langle R|L \rangle =
(-1)^{N/2}2^{1-N/2}
\label{Rf4a}
\end{gather}
and $x$ is the overlap of the two dimer states.  On traversing the MG
point from left to right, the ground state changes from $|+ \rangle
$($|- \rangle $) to $|- \rangle $($|+ \rangle $) for a translationally
invariant system with even (odd) $N/2$ (this is related to Marshall's
sign law \cite{marshall}). It is thus interesting to characterize the
entanglement in the momentum basis $|\pm \rangle $.

\begin{figure}[t] 
\centerline{\resizebox{0.5\textheight}{!}{\rotatebox{270}
{\includegraphics{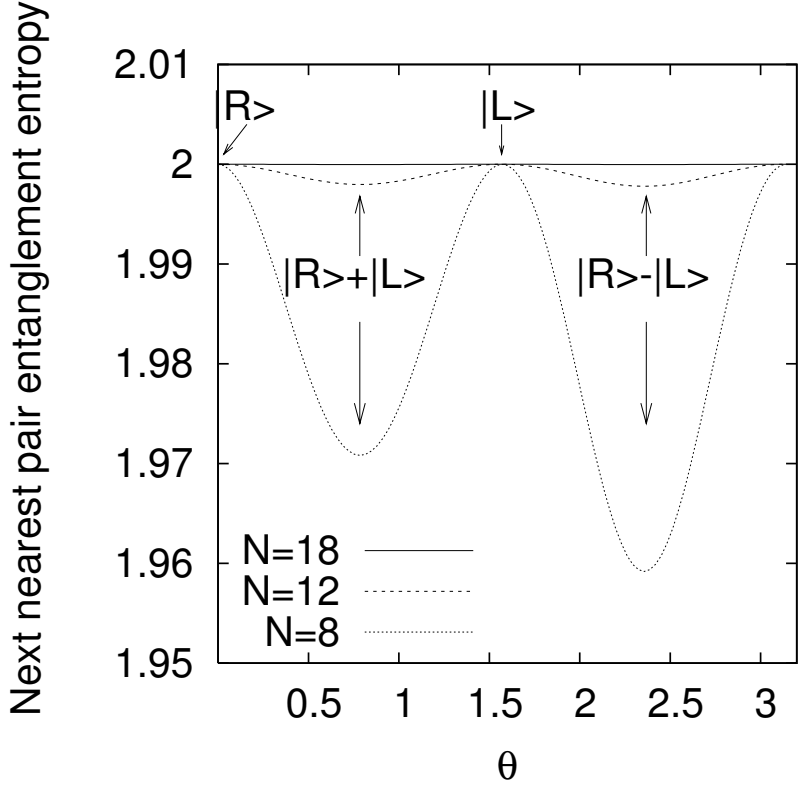}}}}
\caption{Entropy of a pair of n.n.n.  spins in the ground state manifold
  at the MG point}
\label{pairentropy}
\end{figure}

First, consider the entanglement of two nearest neighbour spins, say
$1 $ and $2$ (the same end result holds for all nearest neighbour
spins in these states, due to translational invariance). The parameter
$p$ could be determined by calculating the corresponding correlation
functions. Equivalently, we calculate the form of the state
$\rho_{(1,2)}$ of these two spins explicitly. Notice that the two
spins are bound into a singlet in the state $|R \rangle $, while they
belong to different singlets in the state $|L \rangle$ thus yielding a
maximally mixed reduced state. Hence
\begin{gather}
  \rho_{(1,2)}^{(\pm)}= \text{Tr}' |\pm \rangle \langle \pm| =
  \frac{1}{\Omega_{ \pm}} \text{Tr}'(|R \rangle \langle R| \pm |R \rangle
  \langle L| \pm |L \rangle \langle R| + |L \rangle \langle L|)=
  \nonumber \\ =\frac{1}{\Omega_{ \pm}} ((1\pm 2x)
  |\Psi^{-}\rangle\langle \Psi^{-}| + \openone/4).
\label{Rg4a}
\end{gather}
The entanglement between  two neighbouring spins in the symmetric and
antisymmetric states is determined by the parameters
\begin{gather}
p_{\pm}=  \frac{1\pm 2x}{2 (1\pm x)}.
\label{Rh4a}
\end{gather}
For $N>6$, both states are entangled ($p_{\pm}>1/3$), and so the
difference in the concurrence between them can be calculated
from Eq.(\ref{Rc4a})
\begin{gather}
\Delta  C = C(\rho_{12}^{(+)}) - C(\rho_{12}^{(-)}) = \frac{3}{2}
(p_{+} - p_{-}) = \frac{3 x}{2(1-x^{2})}.
\label{Rj4a}
\end{gather}
The absolute value of this expression gives the ``jump'' in the
nearest neighbour concurrence on traversing the MG point from left to
right (the sign of this difference depends on N, which is consistent
with the ground states in the vicinity of the MG point.). This
quantity has been proposed to be an indicator of the MG point in the
concurrence diagram in Ref. \onlinecite{qian}. However, for large $N$,
$\Delta C$ approaches zero exponentially. Additionally, two spins that
are not nearest neighbours are not entangled with each other, since
the correlation function drops rapidly with distance. Thus a more
general measure considered in \cite{qian}, {\it viz.} the total
concurrence being the sum of concurrences of all pairs contains only
one non-zero contribution coming from nearest neighbours and hence
cannot detect the MG point.

The question thus arises concerning other indicators of this special point.
For the infinite system, a simple candidate could be the dimer order
parameter given by the difference
\begin{gather}
  d= \frac{1}{N}\Bigg|\Big(\sum_{i}^{}\langle \bm{S}_{i} \cdot
  \bm{S}_{i+1}\rangle - \langle \bm{S}_{i+1} \cdot \bm{S}_{i+2} \rangle
  \Big)\Bigg|.
\label{Rp4a}
\end{gather}

At the MG point, $d$ takes the value $3/8$. Considering the notion of
dimerization, one could naively assume that this is the largest
possible value as it corresponds to exact dimers. However, the states
$|R\rangle,|L \rangle $ are not eigenstates of the dimer operator and as such
$d$ cannot take on extremal values for these states. Physically, one
could expect that high but not perfect antiferromagnetic correlations
on one bond supplemented by slightly ferromagnetic correlations on the
other could create a larger value of the dimerization. Using DMRG
techniques, it has been shown that indeed the maximum dimerization
does not occur at the MG point but at $J_{2} \approx 0.5781$
\cite{White}. Additionally, this quantity depends largely on the
working basis: for  translationally invariant states like $|\pm
\rangle $, the value of this parameter is always zero.  Thus the dimer
order parameter also does not distinguish the MG point satisfactorily.

The entanglement entropy of a pair of next nearest neighbour spins is
a much more appropriate quantity that distinguishes the MG point.  Due
to degeneracy at the MG point, the general form of the ground state is
\begin{gather}
  |\Psi_{g}\rangle = \frac{1}{\sqrt{1+x \sin \theta }}\Big(\cos
  \frac{\theta}{2} |R \rangle + \sin \frac{\theta}{2} |L \rangle \Big).
\label{Rm4a}
\end{gather}
The parameter $p_{i,i+2}$ for a pair of n.n.n spins in the above state
is given by 
\begin{gather}
p_{i,i+2} = -\frac{ x \sin \theta }{1+x \sin \theta } .
\label{Rn4a}
\end{gather}

\begin{figure}[t] 
\centerline{\resizebox{0.5\textheight}{!}{\rotatebox{270}
{\includegraphics{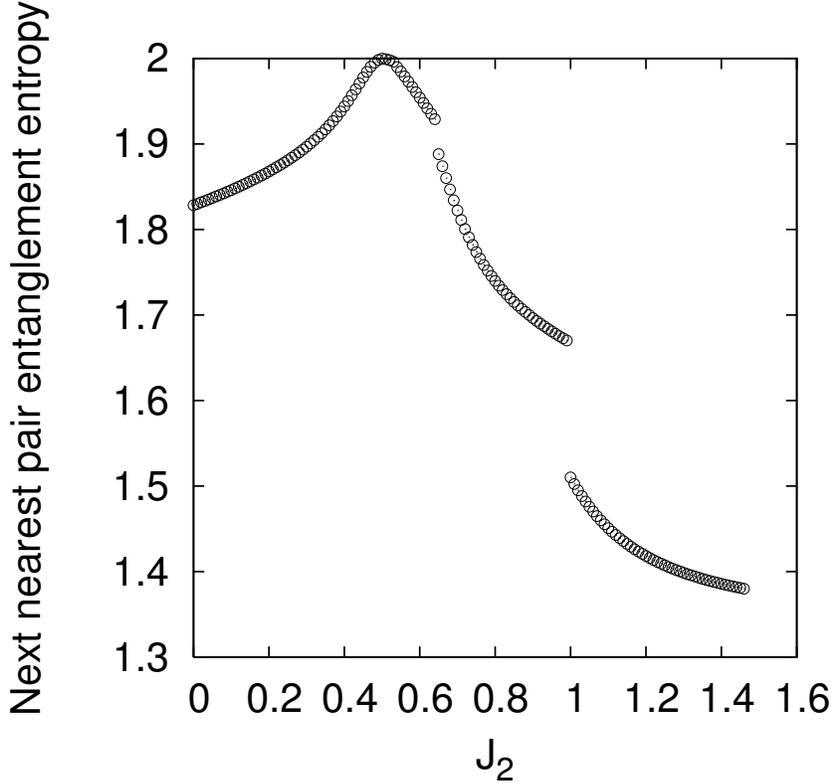}}}} 

\caption{The n.n.n. pair entanglement entropy for
  $N=16$ spins. 
The two jumps which are finite size effects, connected
  to energy level crossings for this system. These discontinuities
  vanish as the system size grows.}
\label{fig7}
\end{figure}

For finite systems, the dimer states $|R\rangle,|L \rangle $ obviously
maximize the entropy of entanglement which is equal to $2$.  Moreover,
as the size of the system increases, the dependence of the entropy on
$\theta $ flattens out to the maximal possible value exponentially
fast, since $x \rightarrow 0$ (see Fig. \ref{pairentropy}). This
further justifies the choice of this quantity as a universal indicator
of the MG point. Interestingly, the momentum states $|\pm \rangle $
are distinguished as local minima in the ground state entanglement
diagram.  In the wider range of values of $J_{2}$, the pair entropy of
the MG point is indeed uniquely distinguished (see Fig. \ref{fig7},
for the n.n.n.  entanglement entropy for 16 spins) as the sole maximum
in the diagram.

\section{Concluding remarks}\label{}

We have focussed on the entanglement properties in the Majumdar-Ghosh
model. Based on data from numerical calculations of finite chains (up
to 28 spins), we have discussed the scaling of the entanglement
entropy of blocks of consecutive spins and shown that it can be used
as a tool to identify the quantum critical point of this model. In
contrast with other numerically studied models, the critical behaviour
of the system does not manifest itself directly in the dependence of
``local'' entanglement on the control parameter, for the considered
system sizes. However, the transition from the critical gapless phase
to the non-critical gapped phase appears in the characteristic change
in scaling of the local entanglement measures w.r.t. total system
sizes. Furthermore, we have shown that the Majumdar-Ghosh point of
this model can be identified as a maximum in the dependence of next
nearest neighbour pair entanglement on the control parameter. 

In the end, we would like to add that one can heuristically consider
the entanglement entropy of the lower rail of spins with the upper
rail of the considered system (Fig. \ref{fig1}) as a natural candidate
for distinguishing the phases of this model. Indeed, the dimer phase
is expected to be characterized by enhanced correlations between the
lower and upper rails of spins.  Once again however, for the system
sizes considered there is no ``characteristic change'' in the
dependence of this quantity on the control parameter. The results have
not been provided in this Article, since they resemble the results
presented in Fig \ref{fig7}. At the MG point, again the ``momentum''
eigenstates reside in local minima, as in Fig. \ref{pairentropy}.

P.T. acknowledges support from Deutsche Forschungsgemeinschaft
(Project No. 436POL 17/17/05). Three of us (R.W.C, P.T. and A.W.)
would like to thank the State Committee for Scientific Research for
financial support under Grant No. 1 P03B 014 30.

\end{document}